\documentclass[showpacs,nofootinbib,tightenlines]{revtex4}   
\usepackage[colorlinks,urlcolor=blue,linkcolor=blue,citecolor=blue]{hyperref} %
\usepackage{epsfig,graphicx,amsmath,amssymb,float,color,xspace,url}

\begin{document}

\title{\large{Resonance contributions $\phi(1020, 1680)\to K\bar K$ for the three-body decays $B\to K\bar K h$} }

\author{Ying-Ying Fan$^1$}
\author{Wen-Fei Wang$^2$}\email{wfwang@sxu.edu.cn}

\affiliation{$^{1}$College of Physics and Electronic Engineering, Xinyang Normal University, Xinyang 464000, China\vspace{-0.2cm}}
\affiliation{$^{2}$Institute of Theoretical Physics, Shanxi University, Taiyuan, Shanxi 030006, China \vspace{0.2cm}}
                   
\date{\today}
\begin{abstract}
We study the contributions for the $K^+K^-$ and $K^0\bar K^0$ originated from the intermediate states $\phi(1020)$ 
and $\phi(1680)$ in the charmless three-body decays $B\to K\bar K h$, with $h=(\pi, K)$, in the perturbative QCD 
approach. The subprocesses $\phi(1020,1680)\to K\bar K$ are introduced into the distribution amplitudes of $K\bar K$ 
system via the kaon electromagnetic form factors with the coefficients taken from the fitted results. The predictions 
of the branching fractions for the decays $B\to\phi(1680)h$ with the intermediate state $\phi(1680)$ decays into $K^+K^-$ or 
$K^0\bar K^0$ are about $6\%$-$8\%$ of the corresponding results for the quasi-two-body decays 
$B\to\phi(1020)h \to K^+ K^- h$ in this work.
\end{abstract}
   
\maketitle
 
\section{INTRODUCTION}\label{sec-int}

Charmless three-body hadronic $B$ meson decays are very important for us to test the Standard Model and to explore the 
Quantum Chromodynamics (QCD). The decay amplitudes of these three-body processes are always described as the coherent sum 
of the resonant and nonresonant contributions in the isobar formalism~\cite{pr135-B551,pr166-1731,prd11-3165}, although isobar 
model violates the unitarity and needs improvement~\cite{PDG-resonance}.  The resonance contributions, 
which are related to the low energy scalar, vector and tensor intermediate states and are associated with the 
various subprocesses of the three-body decays, could be isolated from the total decay amplitudes and studied in the 
quasi-two-body framework~\cite{plb763-29,1605-03889,prd96-113003}. 
The studies of the quasi-two-body decays could also help us to investigate the properties of different resonances and will lead us 
to understand the relationship among the different three-body processes with the same intermediate state.

In addition to the contributions from the $S$-wave intermediate state $f_0(980)$ and the $D$-wave resonance $f^\prime_2(1525)$, 
etc., the $K\bar K$ in the charmless three-body decays $B\to K\bar K h$, with $h$ is pion or kaon, have the contributions from 
the $P$-wave resonances $\rho(770)$, $\omega(782)$, $\phi(1020)$ and their excited states~\cite{prd96-113003}.
The contributions from the resonance $\rho(1450)^0$ and from the tails of the Breit-Wigner (BW) formula~\cite{BW-model} for the 
intermediate states $\rho(770)$ and $\omega(782)$ for $K^+K^-$ in the three-body decays $B^\pm \to K^+K^-\pi^\pm$ have been 
discussed in Ref.~\cite{2004-09027}. In this work, we shall focus on the quasi-two-body decays $B\to\phi(1020,1680)h \to K\bar K h$
within the perturbative QCD (PQCD) approach~\cite{plb504-6,prd63-054008,prd63-074009,ppnp51-85}, with $K\bar K$ is 
the $K^+K^-$ or $K^0\bar K^0$ in the final state. One should note that the $K^0\bar K^0$ which comes from the $P$-wave 
intermediate states could form the $K^0_S$ plus $K^0_L$ but can not generate the $K^0_S$ pair in the final state because of the  
Bose-Einstein statistics. We need to stress that the rescattering effects~\cite{prd71-074016,1512-09284,epjc78-897} in the final 
states were found have important contributions  for the three-body $B$ decays~\cite{prl123-231802}, which would be investigated 
in a subsequent work.

The parameters such as mass and decay width for $\phi(1020)$, the ground state of $s\bar s$, have been measured quite well with the 
processes $e^+e^-\to K^+K^-(\gamma)$ and $e^+e^-\to K^0_SK^0_L$~\cite{plb779-64,prd94-112006,plb760-314,prd88-032013,
plb695-412,plb669-217,prd63-072002}. The $K^+K^-$ and $K^0\bar K^0$ branching fractions for $\phi(1020)$ are consistent with 
the masses dependence in the two-body breakup momentum for the charged and neutral kaon as expected from a $P$-wave 
decay~\cite{prc89-055208}. The structure-dependent radiative corrections to the $\phi(1020)$ decays into $K^+K^-$ and 
$K^0_SK^0_L$ can be found in~\cite{prd78-077301}. The $2^3S_1$ $s\bar s$ state $\phi(1680)$ was discovered in the processes 
of $e^+e^-\to K^0_SK^\pm\pi^\mp$~\cite{pl112B-178},  with the decay dominant into $KK^*(892)$~\cite{pl118B-221,PDG-2018}.
The $K\bar K$ channel for $\phi(1680)$ was found to be about $7\%$ of the $KK^*(892)$ for the branching 
fraction~\cite{pl118B-221}. In Ref.~\cite{jhep1708-037}, the contribution from the subprocess $\phi(1680) \to K^+ K^-$ for 
the three-body decay $B^0_s\to J/\psi K^+K^-$ was found to be $(4.0\pm0.3\pm0.3)\%$ of the total branching fraction by 
LHCb Collaboration recently, which is about $6\%$ of the contribution from $\phi(1020) \to K^+ K^-$ in the same decay channel.
The detailed discussions of the general aspects for $\phi(1680)$ can be found in Ref.~\cite{prd68-054014}.
The $1^{--}$ resonance $\phi(2175)$ was found by BaBar Collaboration~\cite{prd74-091103} and confirmed by different 
experiments~\cite{prd76-012008,prd77-092002,prl100-102003,prd80-031101,prd86-012008,prd91-052017}. In view of its ambiguous 
nature~\cite{prd100-034012}, we shall leave the possible subprocess $\phi(2175)\to K\bar K$ to the future studies.

\begin{figure*}[t]
  \begin{center}
    \includegraphics[width=0.8\textwidth]{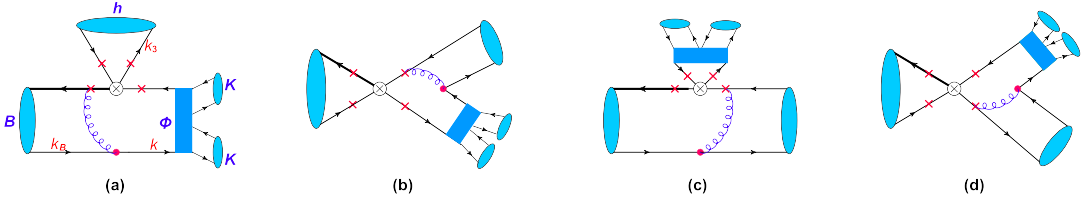}
  \end{center}
  \vspace{-0.2cm}
\caption{Typical Feynman diagrams for $B\to \phi(1020, 1680)h\to KK h$ decays. 
               The $\times$ denotes the possible attachments for hard gluons, the symbol $\phi$ and the rectangle 
               represents the resonances $\phi(1020)$ and $\phi(1680)$. The symbol $\otimes$ is the weak vertex, $B$, $K$ and
               $h$ stand for the $B^+, B^0, B^0_s$, the final states $K^\pm, K^0, \bar K^0$ and the bachelor state pion or kaon, 
               respectively,  and $k_B, k$ and $k_3$ are the momenta for the spectator quarks.
              }
\label{fig-feyndiag}
\end{figure*}

The intermediate states of the quasi-two-body decays $B\to\phi(1020,1680)h \to K\bar K h$ are generated in the hadronization of the quark-antiquark pair $s\bar s$ as demonstrated in the Fig.~\ref{fig-feyndiag}, in which the factorizable and nonfactorizable diagrams 
have been merged for the sake of simplicity, symbol $B$ in the diagrams stands for the mesons $B^+, B^0$ and $B^0_s$, and the 
inclusion of charge-conjugate processes throughout this work is implied. The subprocesses $\phi(1020,1680)\to K\bar K$ which can 
not be calculated in the PQCD approach, will be introduced into the distribution amplitudes of the $K\bar K$ system by the vector 
meson dominance kaon electromagnetic form factor. The PQCD approach has been adopted in Refs.~\cite{plb561-258,prd70-054006,
prd89-074031,prd91-094024} for the tree-body $B$ decays, and the quasi-two-body framework based on PQCD has been discussed 
in detail in~\cite{plb763-29} which has been followed by the works~\cite{2005-02097,2003-03754,jhep2003-162,epjc79-37,
prd96-036014,prd95-056008} for the charmless quasi-two-body $B$ meson decays recently. Parallel analyses of the three-body $B$ 
decays with the QCD factorization (QCDF) can be found in Refs.~\cite{plb622-207,prd74-114009,prd79-094005,prd81-094033,
prd72-094003,prd76-094006,prd88-114014,prd89-074025,prd89-094007,prd94-094015,npb899-247,2007-02558}, and the relevant 
works within the symmetries are referred to Refs~\cite{plb564-90,prd72-075013,prd72-094031,prd84-056002,plb727-136,plb726-337,
prd89-074043,plb728-579,ijmpa29-1450011,prd91-014029}.

This paper is organized as follows. In Sec.~\ref{sec-fra}, we give a brief review of the vector time-like form factors for kaon, 
we present the $P$-wave $K\bar K$ system distribution amplitudes and the differential branching fractions. 
In Sec.~\ref{sec-res}, we provide numerical results for the concerned decay processes and give some necessary discussions. 
Summary of this work is presented in Sec.~\ref{sec-con}. The relevant quasi-two-body decay amplitudes are collected in 
the Appendix.

\section{FRAMEWORK} \label{sec-fra}

In the light-cone coordinates, with the mass $m_B$, the momenta $p_B$ for the $B$ meson and $k_B$ for its light spectator 
quark are written as 
\begin{eqnarray}
p_B=\frac{m_B}{\sqrt2}(1,1,0_{\rm T}),\quad  k_B=\left(\frac{m_B}{\sqrt2}x_B, 0, k_{B{\rm T}}\right)
\end{eqnarray}
in the rest frame of $B$ meson. For the kaon pair generated from the intermediate state $\phi(1020)$ or $\phi(1680)$ 
by the strong interaction, we have its momentum $p=\frac{m_B}{\sqrt 2}(\zeta, 1, 0_{\rm T})$ and the longitudinal 
polarization vector $\epsilon_L=\frac{1}{\sqrt 2}(-\sqrt\zeta, 1/\sqrt\zeta, 0_{\rm T})$, with the variable $\zeta=s/m^2_B$ 
and the invariant mass square $s=m^2_{KK}\equiv p^2$.
The spectator quark comes out from $B$ meson and goes into resonance in the hadronization as shown in 
Fig.~\ref{fig-feyndiag}~(a) has the momentum $k=(0, \frac{m_B}{\sqrt 2}z, k_{\rm T})$. 
For the bachelor final state pion or kaon and its spectator quark, we define their momenta $p_3$ and $k_3$ as
\begin{eqnarray}
p_3=\frac{m_B}{\sqrt2}(1-\zeta, 0, 0_{\rm T}),\quad
k_3=\left(\frac{m_B}{\sqrt2}(1-\zeta)x_3, 0, k_{3{\rm T}}\right)\!.
\end{eqnarray}
The $x_B$, $z$ and $x_3$ above, which run from zero to one in the numerical calculation, are the momentum fractions for $B$ 
meson, intermediate state and the bachelor final state, respectively.

The vector time-like form factors $F_{K^+}(s)$ and $F_{K^0}(s)$ for the charged and neutral kaons are related to the electromagnetic 
form factors for $K^+$ and $K^0$, respectively, which are defined as~\cite{epjc39-41}
\begin{eqnarray} 
  \langle K^+(p_1) K^-(p_2) | j^{em}_\mu | 0 \rangle &=& (p_1-p_2)_\mu \,F_{K^+}(s), \\
  \langle K^0(p_1)\bar K^0(p_2) | j^{em}_\mu | 0 \rangle &=& (p_1-p_2)_\mu\,F_{K^0}(s), 
\end{eqnarray}
with the squared invariant mass $s=(p_1+p_2)^2$, the constraints $F_{K^+}(0)=1$ and $F_{K^0}(0)=0$, and the electromagnetic 
current $j^{em}_\mu=\frac23\bar u\gamma_\mu u-\frac13\bar d\gamma_\mu d -\frac13\bar s\gamma_\mu s$ carried by the light 
quarks $u, d$ and $s$~\cite{npb250-517}. The form factors $F_{K^+}$ and $F_{K^0}$ can be separated into the isospin $I=1$ and 
$I=0$ components as $F_{K^{+(0)}}=F_{K^{+(0)}}^{I=1} + F_{K^{+(0)}}^{I=0}$, with the $F_{K^+}^{I=0}=F_{K^0}^{I=0}$ and 
$F_{K^+}^{I=1}=-F_{K^0}^{I=1}$, and 
$\langle K^+(p_1) \bar{K}^0(p_2) | \bar u \gamma_\mu d | 0 \rangle =(p_1-p_2)_\mu 2F_{K^+}^{I=1}(s)$~\cite{epjc39-41,prd96-113003}.

With the BW formula for the resonances $\omega$ and $\phi$ and the Gounaris-Sakurai (GS) model~\cite{prl21-244} for $\rho$, 
we have the electromagnetic form factors~\cite{epjc39-41,prd81-094014,jetp129-386}
\begin{eqnarray} 
  F_{K^+}(s)&=&+\frac12\sum_{i=\rho,\rho^\prime,...} c^K_i {\rm GS}_i(s) 
                    +\frac16\sum_{i=\omega,\omega^\prime,...} c^K_i {\rm BW}_i(s) 
                   +\frac13\sum_{i=\phi,\phi^\prime,..} c^K_i {\rm BW}_i(s),  
  \label{def-F-K+}   \\
  F_{K^0}(s)&=&-\frac12\sum_{i=\rho,\rho^\prime,...} c^K_i {\rm GS}_i(s) 
                    +\frac16\sum_{i=\omega,\omega^\prime,...} c^K_i {\rm BW}_i(s) 
                   +\frac13\sum_{i=\phi,\phi^\prime,..} c^K_i {\rm BW}_i(s),
  \label{def-F-K0}
\end{eqnarray} 
where the $\sum$ means the summation for the resonances $\rho, \omega$ or $\phi$ and their corresponding excited states, 
$c^K_i$ is proportional to the coupling constant $g_{iK\bar K}$ and the coefficients have the constraints~\cite{jetp129-386}
\begin{eqnarray} 
\sum_{i=\rho,\rho^\prime,...} c^K_i  =1, \qquad
\frac13\sum_{i=\omega,\omega^\prime,...} c^K_i +\frac23\sum_{i=\phi,\phi^\prime,..} c^K_i =1
\end{eqnarray} 
to provide the proper normalizations of the form factors $F_{K^+}(0)=1$ and $F_{K^0}(0)=0$.
One should that the possibility of $SU(3)$ violations are allowed which could and will become manifest in differences between the 
fitted normalization coefficients~\cite{epjc39-41}.
The explicit expressions and auxiliary functions for BW and GS are referred to Refs.~\cite{prd86-032013,prl21-244}. 

Phenomenologically, the vector time-like form factor for kaon can also be defined by~\cite{prd88-114014}
\begin{eqnarray} 
  \langle K^+(p_1) K^-(p_2) |\bar q\gamma_\mu q | 0 \rangle &=& (p_1-p_2)_\mu \,F_{K^+K^-}^{q}(s), \\
  \langle K^0(p_1)\bar K^0(p_2) |\bar q\gamma_\mu q | 0 \rangle &=& (p_1-p_2)_\mu\,F_{K^0\bar K^0}^{q}(s).
\end{eqnarray}
When considering only the resonance contributions, we have
\begin{eqnarray} 
  F_{K^+K^-}^{u}&=&F_{K^0\bar K^0}^{d}= F_\rho+3 F_\omega, \\
  F_{K^+K^-}^{d}&=&F_{K^0\bar K^0}^{u}=-F_\rho+3 F_\omega, \\
  F_{K^+K^-}^{s}&=&F_{K^0\bar K^0}^{s}=-3 F_\phi.
  \label{def-F-uds}
\end{eqnarray}
Then the electromagnetic form factors can also be expressed by $F_{K^+}=F_\rho+F_\omega+F_\phi$ and 
$F_{K^0}=-F_\rho+F_\omega+F_\phi$~\cite{prd88-114014} for the resonance components. 
The expressions for $F_\rho, F_\omega$ and $F_\phi$ can be found in~\cite{prd67-034012,prd88-114014}. 
It's easy to check that 
\begin{eqnarray} 
    F_\phi=\frac13\sum_\phi c^K_\phi {\rm BW}_\phi(s), \quad 
    F_{K^+K^-}^{s}=F_{K^0\bar K^0}^{s}=-\sum_\phi c^K_\phi {\rm BW}_\phi(s).
\end{eqnarray}
We concern only the $\phi$ component of the vector kaon time-like form factors in this work. Rather, for simplicity, we 
employ $F_K$ to stands for $F_{K^+K^-}^{s}$ and $F_{K^0\bar K^0}^{s}$ in the following discussions. 

For the subprocesses $\phi(1020,1680)\to K\bar K$, the $P$-wave $K\bar K$ system distribution amplitudes are organized 
into~\cite{2004-09027,prd76-074018}
\begin{eqnarray} 
  \phi^{P\text{-wave}}_{K\bar K}(z,s)=\frac{-1}{\sqrt{2N_c}}
      \left[\sqrt{s}\,{\epsilon\hspace{-1.5truemm}/}\!_L\phi^0(z,s) +  {\epsilon\hspace{-1.5truemm}/}\!_L {p\hspace{-1.7truemm}/} \phi^t(z,s)   
             +\sqrt s \phi^s(z,s)  \right],
\end{eqnarray}
with the momentum $p=p_1+p_2$.   
We have the distribution amplitudes   
\begin{eqnarray}
   \phi^{0}(z,s)&=&\frac{3F_K(s)}{\sqrt{2N_c}} z(1-z)\left[1+a_2^{\phi} C^{3/2}_2(1-2z) \right],\label{def-DA-0}\\
   \phi^{s}(z,s)&=&\frac{3F^s_K(s)}{2\sqrt{2N_c}}(1-2z), \label{def-DA-s}\\    
   \phi^{t}(z,s)&=&\frac{3F^t_K(s)}{2\sqrt{2N_c}}(1-2z)^2,\label{def-DA-t}  
\end{eqnarray}
with the Gegenbauer polynomial $C^{3/2}_2(\chi)=3\left(5\chi^2-1\right)/2$ and 
$F^{s,t}_K(s)\approx (f^T_{\phi}/f_{\phi})F_K(s)$~\cite{plb763-29} with the ratio $f^T_{\phi}/f_{\phi}=0.75$ at the scale $\mu=2$
GeV~\cite{prd78-114509}. The Gegenbauer moment $a_2^{\phi}$ for $\phi^{0}(z,s)$ is the same as it in the distribution amplitudes 
of the light vector meson $\phi$ in~\cite{prd76-074018} for the two-body $B$ meson decays.

The $CP$ averaged differential branching fractions ($\mathcal B$) for the quasi-two-body decays $B\to\phi(1020,1680)h \to K\bar K h$ 
are written as~\cite{prd79-094005,2004-09027,plb791-342}
\begin{eqnarray}
 \frac{d{\mathcal B}}{d\zeta}=\tau_B\frac{q^3_h q^3}{12\pi^3m^5_B}\overline{|{\mathcal A}|^2}\;,
\label{eqn-diff-bra}
\end{eqnarray}
where $\tau_B$ being the $B$ meson mean lifetime. The magnitudes of the momenta $q$ and $q_h$ for the kaon and the bachelor 
$h$ in the rest frame of the resonances $\phi(1020,1680)$ are written as
\begin{eqnarray}
       q&=&\frac{1}{2}\sqrt{s-4m^2_K},  \label{def-q}\\
   q_h&=&\frac{1}{2\sqrt s}\sqrt{\left(m^2_{B}-m_{h}^2\right)^2 -2\left(m^2_{B}+m_{h}^2\right)s+s^2},   \label{def-qh}
\end{eqnarray}
with the mass $m_h$ for the bachelor meson pion or kaon.
The direct $CP$ asymmetry ${\mathcal A}_{CP}$ is defined as
\begin{eqnarray}
{\mathcal A}_{CP}=\frac{{\mathcal B}(\bar B\to \bar f)-{\mathcal B}(B\to f)}{{\mathcal B}(\bar B\to \bar f)+{\mathcal B}(B\to f)}
\end{eqnarray}
The Lorentz invariant decay amplitudes for the quasi-two-body decays $B\to\phi(1020,1680)h \to K\bar K h$ are collected in the Appendix.

\section{RESULTS} \label{sec-res}

In the numerical calculation, we employ the decay constants $f_B=0.189$ GeV and $f_{B_s}=0.231$ GeV~\cite{prd98-074512}, 
the mean lives $\tau_{B^0}=(1.520\pm0.004)\times 10^{-12}$~s, $\tau_{B^+}=(1.638\pm0.004)\times 10^{-12}$~s and 
 $\tau_{B^0_s}=(1.509\pm0.004)\times 10^{-12}$~s~\cite{PDG-2018} for the $B^0, B^+$ and $B^0_s$ mesons, respectively. 
The masses and the decay constants for the relevant particles in this work, the full widths for $\phi(1020)$ and $\phi(1680)$, 
and the Wolfenstein parameters of the Cabbibo-Kobayashi-Maskawa (CKM) matrix are presented in Table~\ref{tab1}.

\begin{table}[thb]  
\begin{center}
\caption{Masses, decay constants, full widths of $\phi(1020)$ and $\phi(1680)$ (in units of GeV) and Wolfenstein 
parameters~\cite{PDG-2018}.}
\label{tab1}
\begin{tabular}{l}\hline\hline 
\;$m_{B^{0}}=5.280 \quad\; m_{B^{\pm}}=5.279 \quad m_{B^0_s}=5.367 \quad m_{\pi^\pm}=0.140 \quad\; m_{\pi^0}=0.135$ \\
\;$m_{K^\pm}=0.494 \quad  m_{K^0}=0.498 \quad\; f_K=0.156\quad\; f_\pi=0.130\;\;\;  m_{\phi(1020)}=1.019 $ \\ 
\;$\Gamma_{\phi(1020)}=0.00425 \quad\;\;    m_{\phi(1680)}=1.680\pm0.020 \quad\;\;\; \Gamma_{\phi(1680)}=0.150\pm0.050$  \\
\;$\lambda=0.22453\pm 0.00044  \;\;  A=0.836\pm0.015  \;\;  \bar{\rho} = 0.122^{+0.018}_{-0.017}  \;\;\; 
    \bar{\eta}= 0.355^{+0.012}_{-0.011} $\\ 
\hline\hline  
\end{tabular}
\end{center}
\end{table}

The coefficients $c^K_{\phi(1020)}$ and $c^K_{\phi(1680)}$ in the electromagnetic form factors $F_{K^+}$ and $F_{K^0}$, the 
Eqs.~(\ref{def-F-K+})-(\ref {def-F-K0}), have been fitted to the data in Refs.~\cite{epjc39-41,prd81-094014,jetp129-386}.
The results of the constrained and unconstrained fits in~\cite{epjc39-41,prd81-094014} and the results of the Model I and Model II 
in~\cite{jetp129-386} for $c^K_{\phi(1020)}$ agree with each other. But the fit results for $c^K_{\phi(1680)}$ 
are quite different in Refs.~\cite{epjc39-41,prd81-094014,jetp129-386}, with the results $-0.018\mp0.006$ $(0.001\mp0.007)$ and 
$0.0042\pm0.0015$ $(0.0136\pm0.0024)$ of the constrained (unconstrained) fits in~\cite{epjc39-41} and~\cite{prd81-094014}, 
respectively, and $-0.117\pm0.020$ $(-0.150\pm0.009)$ for the Model I (II) in~\cite{jetp129-386}. While one can find that the coefficient 
$c_{\rho(1450)}$ for the pion electromagnetic form factor $F_\pi$ in Refs.~\cite{prd86-032013,zpc76-15,prd61-112002,pr421-191,
prd78-072006} by different collaborations are consistent with each other. 
By refer to the discussions in~\cite{epjc39-41} for hadronic invariant amplitudes for $\rho\to\pi^+\pi^-$ and $\phi\to K^+K^-$,
one could obtain the relation $|g_{\phi(1680)KK}|\approx |g_{\rho(1450)\pi\pi}|/\sqrt2$ within $SU(3)$ symmetry.
With the relations
\begin{eqnarray} 
|c_{\rho(1450)}|\approx\frac{f_{\rho(1450)}|g_{\rho(1450)\pi\pi}|}{\sqrt2 m_{\rho(1450)}}, \qquad
|c^K_{\phi(1680)}|\approx\frac{f_{\phi(1680)}|g_{\phi(1680)KK}|}{ m_{\phi(1680)}}, 
\label{eqs-crho+ck}
\end{eqnarray}  
and the result $|c_{\rho(1450)}|=0.178\footnote{One should note the different definitions for the coefficient $c_{\rho}$ in~\cite{prd86-032013} and~\cite{epjc39-41}.}$
from~\cite{prd86-032013}, it's easy to get $|c^K_{\phi(1680)}|\approx0.160$ supposing the decay constants 
$f_{\rho(1450)}/f_{\phi(1680)}\approx f_{\rho(770)}/f_{\phi(1020)}$.  
With the partial width ratio~\cite{pl118B-221} 
\begin{eqnarray} 
\frac{\Gamma{(KK)}}{\Gamma{(KK^\ast(892))}}\approx0.073, 
\end{eqnarray}
and the rough branching ratio ${\mathcal B}_{KK^\ast(892)}\approx0.7$~\cite{prd86-012008,plb798-134946} for the resonance 
$\phi(1680)$, one could estimate $|c^K_{\phi(1680)}|\approx0.092$. While with the 
decay widths $19.8\pm4.3$ MeV in~\cite{prd96-054033} and $17$ MeV in~\cite{plb744-1} for $\phi(1680)\to K\bar K$, we 
estimate the coefficient $|c^K_{\phi(1680)}|$ at about $0.130$-$0.162$. 
In view of our estimated values, 
we employ the fitted result $c^K_{\phi(1680)}=-0.150\pm0.009$~\cite{jetp129-386} in our numerical calculation in this work. 
As for the coefficient $c^K_{\phi(1020)}$ of the electromagnetic form factors $F_{K^+}$ and $F_{K^0}$, we adopt its 
fitted value $1.038$ in the Model II in Ref.~\cite{jetp129-386}.

Utilizing the differential branching fraction the Eq.~(\ref{eqn-diff-bra}) and the decay amplitudes collected in Appendix \ref{sec-app}, 
we obtain the concerned direct $CP$ asymmetries and the $CP$ averaged branching fractions for the quasi-two-body decays 
$B\to\phi(1020)h \to K\bar K h$ in Table~\ref{Res-phi1020} and $B\to\phi(1680)h \to K^+ K^- h$ in Table~\ref{Res-phi1680}.
Only the modes $B^+\to \phi(1020,1680) K^+$ and $B_s^0 \to\phi(1020,1680) \pi^0$ with $\phi(1020,1680)$ decay into 
$K^+K^-$ or $K^0\bar K^0$, which contain the contributions from the current-current operators of the weak effective 
Hamiltonian~\cite{rmp68-1125}, have the direct $CP$ asymmetries in Tables~\ref{Res-phi1020},~\ref{Res-phi1680}.
The first error of these results in Tables~\ref{Res-phi1020},~\ref{Res-phi1680} comes from the uncertainty of the shape parameters
$\omega_B=0.40\pm0.04$ for $B^+$ and $B^{0}$ and $\omega_{B}=0.50\pm0.05$ for $B^0_s$, the second error is induced  
by the chiral scale parameters $m^\pi_0=1.40\pm0.10$ GeV and $m^K_0=1.60\pm0.10$ GeV, which are defined using the meson 
masses and the component quark masses as $m^{\pi(K)}_0=\frac{m_{\pi(K)}}{m_q+m_{q^\prime}}$, 
and the Gegenbauer moment $a^{\pi,K}_2=0.25\pm0.15$ for $\pi$ and $K$ as in~\cite{prd86-114025}, the third one is contributed by the Gegenbauer moment $a_2^{\phi}=0.18\pm0.08$~\cite{prd76-074018} and 
the fourth error in Table~\ref{Res-phi1680} comes from the variation of the coefficient $c^K_{\phi(1680)}$ of the form factor $F_K$, 
which will not change the direct $CP$ asymmetries. There are other errors come from the uncertainties of the masses and the decay 
constants of the initial and final states, the other parameters in the distribution amplitudes of the bachelor pion or kaon, the Wolfenstein 
parameters of the CKM matrix, etc. are small and have been neglected.

\begin{table}[thb]   
\begin{center}
\caption{PQCD predictions of the $CP$ averaged branching fractions and the direct $CP$ asymmetries 
              for the $B\to\phi(1020)h \to K\bar K h$ decays.}
\label{Res-phi1020}   
\begin{tabular}{l c l} \hline\hline
   \quad Decay modes       &    ~    &  \quad\; Quasi-two-body results               \\
\hline  %
  $B^+\to  \phi(1020) K^+ \!\to K^+K^- K^+$    &${\mathcal B}(10^{-6})$\;
      & $4.03\pm0.67(\omega_B)\pm0.49(m^K_0{+}a^K_2)\pm0.15(a^\phi_2)$    \\
                &  \;${\mathcal A}_{CP}(\%)$\;\;
      & $-1.31\!\pm0.91(\omega_B)\!\pm2.63(m^K_0{+}a^K_2)\!\pm0.55(a^\phi_2)$   \\
  $B^+\to  \phi(1020)\pi^+ \,\to K^+K^- \pi^+$    &${\mathcal B}(10^{-9})$\;
      & $3.58\pm1.17(\omega_B)\pm1.87(m^\pi_0{+}a^\pi_2)\pm0.34(a^\phi_2)$    \\
  $B^0 \;\to \phi(1020) K^0\to K^+K^- K^0$\    &${\mathcal B}(10^{-6})$\;
      & $3.62\pm0.64(\omega_B)\pm0.59(m^K_0{+}a^K_2)\pm0.19(a^\phi_2)$    \\     
  $B^0 \;\to \phi(1020) \pi^0 \;\to K^+K^- \pi^0$\    &${\mathcal B}(10^{-9})$\;
      & $1.74\pm0.53(\omega_B)\pm0.91(m^\pi_0{+}a^\pi_2)\pm0.14(a^\phi_2)$    \\       
  $B_s^0 \;\to  \phi(1020) \bar K^0\to K^+K^- \bar K^0$ &${\mathcal B}(10^{-8})$\;
      & $8.34\pm0.48(\omega_B)\pm0.94(m^K_0{+}a^K_2)\pm2.07(a^\phi_2)$    \\         
  $B_s^0 \;\to  \phi(1020) \pi^0 \;\to K^+K^- \pi^0$ &${\mathcal B}(10^{-8})$\;
      & $9.11\pm2.03(\omega_B)\pm0.14(m^\pi_0{+}a^\pi_2)\pm0.61(a^\phi_2)$    \\        
                &  \;${\mathcal A}_{CP}(\%)$\;\;  
      & $10.58\!\pm1.89(\omega_B)\!\pm2.01(m^\pi_0{+}a^\pi_2)\!\pm0.84(a^\phi_2)$   \\
\hline           
  $B^+\to  \phi(1020) K^+ \!\to K^0\bar K^0 K^+$    &${\mathcal B}(10^{-6})$\;
      & $2.79\pm0.46(\omega_B)\pm0.34(m^K_0{+}a^K_2)\pm0.11(a^\phi_2)$    \\
  $B^+\to  \phi(1020)\pi^+ \,\to K^0\bar K^0 \pi^+$    &${\mathcal B}(10^{-9})$\;
      & $2.47\pm0.81(\omega_B)\pm1.30(m^\pi_0{+}a^\pi_2)\pm0.24(a^\phi_2)$    \\
  $B^0 \;\to \phi(1020) K^0\to K^0\bar K^0 K^0$\    &${\mathcal B}(10^{-6})$\;
      & $2.50\pm0.44(\omega_B)\pm0.41(m^K_0{+}a^K_2)\pm0.13(a^\phi_2)$    \\  
  $B^0 \;\to \phi(1020) \pi^0 \;\to K^0\bar K^0 \pi^0$\    &${\mathcal B}(10^{-9})$\;
      & $1.20\pm0.37(\omega_B)\pm0.63(m^\pi_0{+}a^\pi_2)\pm0.10(a^\phi_2)$    \\   
  $B_s^0 \;\to  \phi(1020) \bar K^0\to K^0\bar K^0 \bar K^0$ &${\mathcal B}(10^{-8})$\;
      & $5.76\pm0.33(\omega_B)\pm0.65(m^K_0{+}a^K_2)\pm1.44(a^\phi_2)$    \\        
  $B_s^0 \;\to  \phi(1020) \pi^0 \;\to K^0\bar K^0 \pi^0$ &${\mathcal B}(10^{-8})$\; 
      & $6.30\pm1.40(\omega_B)\pm0.10(m^\pi_0{+}a^\pi_2)\pm0.43(a^\phi_2)$    \\                            
\hline\hline
\end{tabular}
\end{center}
\end{table}

\begin{table}[thb]   
\begin{center}
\caption{PQCD predictions of the $CP$ averaged branching fractions and the direct $CP$ asymmetries 
              for the $B\to\phi(1680)h \to K^+ K^- h$ decays. The decay mode with the subprocess $\phi(1680)\to K^0\bar K^0$ 
              has the same branching fraction and direct $CP$ asymmetry of its corresponding decay with $\phi(1680)\to K^+K^-$.}
\label{Res-phi1680}   
\begin{tabular}{l c l} \hline\hline
   \quad Decay modes       &    ~    &  \quad\; Quasi-two-body results               \\
\hline  %
  $B^+\to  \phi(1680) K^+\!\to K^+K^- K^+$   &${\mathcal B}(10^{-7})$\;
      & $2.51\pm0.35(\omega_B)\pm0.42(m^K_0{+}a^K_2)\pm0.15(a^\phi_2)\pm0.30(c^K_\phi)$     \\
                &  \;${\mathcal A}_{CP}(\%)$\;\;
      & $-1.39\!\pm0.85(\omega_B)\!\pm1.33(m^K_0{+}a^K_2)\!\pm0.67(a^\phi_2)\pm0.00(c^K_\phi)$  \\
  $B^+\to  \phi(1680)\pi^+\,\to K^+K^- \pi^+$    &${\mathcal B}(10^{-10})$\;
      & $2.84\pm0.98(\omega_B)\pm1.73(m^\pi_0{+}a^\pi_2)\pm0.26(a^\phi_2)\pm0.34(c^K_\phi)$    \\
  $B^0\;\to \phi(1680) K^0\to K^+K^- K^0$\    &${\mathcal B}(10^{-7})$\;
      & $2.39\pm0.36(\omega_B)\pm0.41(m^K_0{+}a^K_2)\pm0.18(a^\phi_2)\pm0.29(c^K_\phi)$    \\     
  $B^0\;\to \phi(1680) \pi^0\;\to K^+K^- \pi^0$\    &${\mathcal B}(10^{-10})$\;
      & $1.46\pm0.45(\omega_B)\pm0.73(m^\pi_0{+}a^\pi_2)\pm0.16(a^\phi_2)\pm0.18(c^K_\phi)$    \\       
  $B_s^0\;\to  \phi(1680) \bar K^0\to K^+K^- \bar K^0$ &${\mathcal B}(10^{-9})$\;
      & $6.25\pm0.40(\omega_B)\pm1.05(m^K_0{+}a^K_2)\pm1.40(a^\phi_2)\pm0.75(c^K_\phi)$   \\         
  $B_s^0\;\to  \phi(1680) \pi^0\;\to K^+K^- \pi^0$ &${\mathcal B}(10^{-9})$\;
      & $6.47\pm1.38(\omega_B)\pm0.17(m^\pi_0{+}a^\pi_2)\pm0.31(a^\phi_2)\pm0.78(c^K_\phi)$    \\        
                &  \;${\mathcal A}_{CP}(\%)$\;\;
      & $5.16\pm1.37(\omega_B)\pm1.46(m^\pi_0{+}a^\pi_2)\pm0.38(a^\phi_2)\pm0.00(c^K_\phi)$   \\                          
\hline\hline
\end{tabular}
\end{center}
\end{table}

The two-body branching fractions for $B\to \phi h$ can be extracted from the quasi-two-body predictions with the relation 
\begin{eqnarray}
\Gamma(B\to \phi h\to K\bar K h)\approx\Gamma(B\to \phi h)\cdot{\mathcal B}(\phi \to K\bar K).
\label{def-fac-rel}
\end{eqnarray}
In Ref.~\cite{jhep2003-162}, a parameter $\eta$ was defined to measure the violation of the factorization relation the 
Eq.~(\ref{def-fac-rel}) for the $B\to K_0^{*}(1430) h$ and $B\to K_0^{*}(1430) h\to K\pi h$ decays. 
For the decays $B\to \phi(1020) h$ and $B\to \phi(1020) h\to K\bar K h$ in this work, 
we have the definition
\begin{eqnarray}
 \eta&=&\frac{\Gamma(B\to \phi(1020)h\to K\bar K h)}{\Gamma(B\to \phi(1020)h)\cdot{\mathcal B}(\phi(1020)\to K\bar K)}\nonumber\\
       &\approx&\frac{m^4_{\phi(1020)}}{384\pi^2 m^3_{B} f^2_\phi \hat q^3_h }  
             \frac{1}{{\mathcal B}(\phi(1020)\to K\bar K) }\nonumber\\
       &\times& \int^{(m_{B}-m_h)^2}_{4 m_K^2}\frac{ds}{s^3}\frac{\lambda^{3/2}(m^2_B,s,m^2_h)\lambda^{3/2}(s,m^2_K,m^2_K) }
             {(s-m^2_{\phi(1020)})^2+(m_{\phi(1020)}\Gamma_{\phi(1020)}(s))^2}, \;\;
\label{def-parm-eta}
\end{eqnarray}
where $\lambda(a,b,c)=a^2+b^2+c^2-2ab-2ac-2bc$, the $\hat q_h$ is the expression of Eq.~(\ref{def-qh}) in the rest frame 
of $B$ meson and fixed at $s=m^2_{\phi}$. As an example, we have $\eta\approx1.07$ for the decays $B^0\to \phi(1020) K^0$ 
and $B^0\to \phi(1020) K^0\to K^+K^- K^0$ with the branching fraction 
${\mathcal B}(\phi(1020)\to K^+K^-)=0.492$~\cite{PDG-2018}. It means that the violation of the Eq.~(\ref{def-fac-rel}) is 
small when neglecting the effect of the squared invariant mass $s$ in the decay amplitudes of the quasi-two-body decays. 
As the verification of Eq.~(\ref{def-parm-eta}), we calculate the decay $B^0\to \phi(1020) K^0$ in 
the two-body framework of the PQCD approach 
with the same parameters and obtain its branching fraction ${\mathcal B}(B^0\to \phi(1020) K^0)\approx7.21\times 10^{-6}$, 
which is about $98.0\%$ of the result $7.36\times 10^{-6}$ in Table~\ref{tab-data} extracted with the corresponding quasi-two-body 
result in Table~\ref{Res-phi1020} with the factorization relation.

The comparison of the extracted PQCD predictions with the experimental measurements for the relevant two-body branching 
fractions are shown in Table~\ref{tab-data}. The branching ratio $8.8^{+0.7}_{-0.6}\times 10^{-6}$ for the two-body decay 
$B^+\to  \phi(1020) K^+$, which was averaged from the results in Refs.~\cite{prd85-112010,prl95-031801,prd71-092003,prl86-3718} 
presented BaBar, CDF, Belle and CLEO Collaborations,  is consistent with the prediction $(8.19\pm1.71)\times 10^{-6}$ in this work.
The data ${\mathcal B}=(7.3\pm0.7)\times 10^{-6}$ in~\cite{PDG-2018} averaged from the results in~\cite{prd85-112010,
prl91-201801,prl87-151801,prd69-011102} for the decay $B^0\to  \phi(1020) K^0$  agree well with our prediction 
$(7.36\pm1.81)\times 10^{-6}$ in Table~\ref{tab-data}.  In~\cite{plb728-85}, an upper limit $1.5\times 10^{-7}$ was set by LHCb 
at $90\%$ confidence level for the branching fraction of the decay $B^{\pm}\to\phi \pi^{\pm}$. Very recently, LHCb Collaboration 
presented a fit fraction $(0.3\pm0.1\pm0.1)\%$ of the total branching fraction of $B^\pm \to \pi^\pm K^+K^-$ for the subprocess 
$\phi(1020)\to K^+K^-$ in Ref.~\cite{prl123-231802}, meaning the two-body branching ratio 
${\mathcal B}(B^{\pm}\to\phi(1020) \pi^{\pm})=(3.2\pm1.5)\times 10^{-8}$~\cite{PDG-2020}, which is larger than the corresponding 
prediction in~\ref{tab-data} but both with large uncertainty.  

\begin{table}[H]  
\begin{center}
\caption{Comparison of the extracted predictions with the experimental measurements for the relevant two-body branching 
               fractions. The errors for the predictions have been added in quadrature.}
\label{tab-data}   
\begin{tabular}{l l l l} \hline\hline
   Two-body decays                & \;\quad This work                                    & \;\quad  Data~\cite{PDG-2018}            \\
\hline
  $B^+\to  \phi(1020) K^+$    &\;\;\;  $(8.19\pm1.71)\times 10^{-6}$\;\;   &\;\;\; $8.8^{+0.7}_{-0.6}\times 10^{-6}$  \\
  $B^+\to  \phi(1020) \pi^+$  &\;\;\;  $(7.28\pm4.54)\times 10^{-9}$\;\;    &\;\;\;  $<1.5\times 10^{-7}$                     \\
  $B^0\to  \phi(1020) K^0$    &\;\;\;  $(7.36\pm1.81)\times 10^{-6}$\;\;    &\;\;\;  $(7.3\pm0.7)\times 10^{-6}$          \\
  $B^0\to  \phi(1020) \pi^0$  &\;\;\;  $(3.54\pm2.16)\times 10^{-9}$\;\;    &\;\;\;  $<1.5\times 10^{-7}$                      \\
\hline\hline
\end{tabular}
\end{center}
\end{table}

The $B^+\to \phi(1020) \pi^+$ was studied in~\cite{npb675-333} with its branching ratio at about $5\times 10^{-9}$ within 
QCDF, which agree with our prediction $(7.28\pm4.54)\times 10^{-9}$ within errors. 
The predicted results in Table~\ref{Res-phi1020} for the decays $B_s^0\to \phi(1020) \bar K^0$ and $B_s^0 \to\phi(1020) \pi^0$ 
are consistent with the theoretical results in Refs.~\cite{npb675-333,prd76-074018,prd78-034011,prd80-114026} within errors 
by considering ${\mathcal B}(\phi(1020)\to K^+K^-)=0.492$~\cite{PDG-2018}. The branching ratios for the two-body decays 
$B\to \phi \pi$ were found to be enhanced by the $\omega$-$\phi$ mixing effect in~\cite{prd80-014024}. The $\omega$-$\phi$ 
mixing effect for the quasi-two-body decays $B\to\phi(1020,1680)h \to K\bar K h$ is out of the scope of this work and will be left 
to the future studies. The penguin-dominated two-body decays $B^\pm\to K^\pm \phi(1020)$ and $B^0\to K^0 \phi(1020)$ have 
been studied in Refs.~\cite{prd74-094020,plb521-252,prd64-112002} within PQCD approach with the consistent results with our 
predicted values in Table~\ref{tab-data}. 

\begin{figure}[tbp]   
\centerline{\epsfxsize=8 cm \epsffile{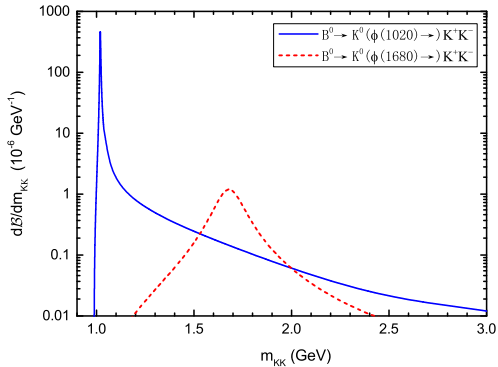}}
     \vspace{-0.2cm}
\caption{Differential branching fractions from the threshold of $K^+K^-$ to $3$ GeV for the 
                 $B^0\to\phi(1020) K^0\to K^+K^- K^0$ and $B^0\to\phi(1680) K^0\to K^+K^- K^0$ decays.}
\label{fig=depbr}
\end{figure}

The predictions for the branching fractions of the decays $B\to\phi(1680)h \to K^+ K^- h$ in Table~\ref{Res-phi1680} are about 
$6\%$-$8\%$ of the corresponding results for $B\to\phi(1020)h \to K^+ K^- h$ in Table~\ref{Res-phi1020}. 
The main portion of these branching fractions for $B\to\phi(1020,1680)h \to K^+ K^- h$ lies in the region around the pole masses 
of the intermediate states $\phi(1020)$ and $\phi(1680)$, which could be concluded from the differential branching fractions 
for the decays $B^0\to\phi(1020) K^0\to K^+K^- K^0$ and $B^0\to\phi(1680) K^0\to K^+K^- K^0$ shown in Fig.~\ref{fig=depbr}.
In Ref.~\cite{jhep1708-037}, the contributions from the subprocesses $\phi(1020) \to K^+ K^-$ and $\phi(1680) \to K^+ K^-$ were 
fitted by LHCb to be $(70.5\pm0.6\pm1.2)\%$ and $(4.0\pm0.3\pm0.3)\%$, respectively, of the total branching fraction for 
the three-body decay $B^0_s\to J/\psi K^+K^-$, implying a ratio at about $0.06$ between the branching fractions of the 
quasi-two-body decays $B^0_s\to J/\psi \phi(1680)\to J/\psi K^+K^-$ and $B^0_s\to J/\psi\phi(1020) \to J/\psi K^+K^-$,
which is consistent with the results $6\%$-$8\%$ in this work for $B\to\phi(1680,1020)h \to K^+ K^- h$.
We need to stress that there will be interference between the contributions from $\phi(1020)$ and $\phi(1680)$ which could 
increase or decrease the total contributions from these two resonances dependents on the phase difference between them. 
According to Fig.~\ref{fig=depbr}, the contribution for $K^+K^-$ from the resonance $\phi(1020)$ is down by a factor of $10$ 
at the peak of $\phi(1680)$ comparing with contribution from $\phi(1680)$, which means that the amplitudes are different by a 
factor about $3$ and the interference allows for a variation between $(1+1/3)^2\approx1.8$ and $(1-1/3)^2\approx0.4$ in the 
region around the pole mass of $\phi(1680)$.

The ratio between branching fractions of the decays  $\phi(1680) \to K^0 \bar K^0$ and $\phi(1680) \to K^+ K^-$ is
close to one because of the coupling constants $g_{\phi(1680) K^0 \bar K^0}=g_{\phi(1680) K^+ K^-}$~\cite{epjc39-41} 
and $m^2_{\phi(1680)}-4m^2_{K^0}  \approx m^2_{\phi(1680)}-4m^2_{K^+}$.
This means that the decay mode with the subprocess $\phi(1680)\to K^0\bar K^0$ has the same branching fraction of its 
corresponding process with $\phi(1680)\to K^+K^-$ for $B\to\phi(1680)h \to K\bar K h$. While for the decays 
$\phi(1020) \to K^0 \bar K^0$ and $\phi(1020) \to K^+ K^-$, one has a ratio $0.66$ between their branching fractions, 
which is consistent with the ratio $0.69$ between the branching fractions in~\cite{PDG-2018} for these two decays, with 
the coupling constants $g_{\phi(1020) K^0 \bar K^0}=g_{\phi(1020) K^+ K^-}$~\cite{plb779-64,epjc39-41}.
The results in Table~\ref{Res-phi1020} for the subprocess $\phi(1020) \to K^0 \bar K^0$ are deduced 
from ${\mathcal B}(\phi(1020) \to K^0 \bar K^0)=34.0\%$~\cite{PDG-2018} along with the results in the same table for the decays
with the subprocess $\phi(1020) \to K^+ K^-$. With the decay amplitude for $B^+\to\phi(1020) K^+ \to K^+K^- K^+$, we calculate 
the branching fraction and direct $CP$ asymmetry for the decay $B^+\to\phi(1020) K^+ \to K^0\bar K^0 K^+$, and obtain 
the central values ${\mathcal B}=2.83\times 10^{-6}$ and ${\mathcal A}_{CP}=-1.25\%$ for it, which are agree well with the results 
in Table~\ref{Res-phi1020} for this process.

\section{Summary} \label{sec-con}
In this work, we studied the contributions for the $K^+K^-$ and $K^0\bar K^0$ which originated from the intermediate 
states $\phi(1020)$ and $\phi(1680)$ in the charmless three-body decays $B\to K\bar K h$ in PQCD approach. 
The subprocesses $\phi(1020,1680)\to K\bar K$ were introduced into the distribution amplitudes of $K\bar K$ system 
via the kaon electromagnetic form factor with the coefficients $c^K_{\phi(1020)}$ and $c^K_{\phi(1680)}$ in which
are adopted from the fitted results. With $c^K_{\phi(1020)}=1.038$ and $c^K_{\phi(1680)}=-0.150\pm0.009$ 
we predicted the branching fractions for the quasi-two-body decays $B\to\phi(1020)h \to K\bar K h$ and 
$B\to\phi(1680)h \to K^+ K^- h$ and the direct $CP$ asymmetries for the decay modes $B^+\to \phi(1020,1680) K^+$ 
and $B_s^0 \to\phi(1020,1680) \pi^0$ with $\phi(1020,1680)$ decay into $K^+K^-$ or $K^0\bar K^0$.

The predictions for the branching fractions of the decays $B\to\phi(1680)h \to K^+ K^- h$ are about $6\%$-$8\%$ 
of the corresponding results for $B\to\phi(1020)h \to K^+ K^- h$ in this work. The branching fraction for the decay 
$\phi(1680) \to K^0 \bar K^0$ is equal to that for $\phi(1680) \to K^+ K^-$, and the decay mode with the subprocess 
$\phi(1680)\to K^0\bar K^0$ has the same branching fraction of its corresponding mode with $\phi(1680)\to K^+K^-$ 
for $B\to\phi(1680)h \to K\bar K h$. We defined a parameter $\eta$ to measure the violation of the factorization relation 
for the decays $B\to\phi h$ and $B\to\phi h \to K\bar K h$ and found the violation is quite small. With the factorization 
relation, we extracted the branching fractions for the two-body decays $B^{0,+}\to  \phi(1020) K^{0,+}$ and 
$B^{0,+}\to  \phi(1020) \pi^{0,+}$. The predictions for the decays $B^0\to  \phi(1020) K^0$ and $B^{+}\to  \phi(1020) K^{+}$ 
are agree with the existing data. And our results for $B^{0,+}\to  \phi(1020) \pi^{0,+}$ consistent with the theoretical results 
in literature.

\begin{acknowledgments}
This work was supported in part by the National Natural Science Foundation of China under Grants No. 11505148, 
No. 11547038 and No. 11575110. Y.Y. Fan was also supported by the Nanhu Scholars Program for Young Scholars of XYNU. 
W.F. Wang thank Ai-Jun Ma for valuable discussions.
\end{acknowledgments}                 

\appendix
\section{DECAY AMPLITUDES} \label{sec-app}

The Lorentz invariant decay amplitude ${\mathcal A}$ for the quasi-two-body processes  $B\to \phi(1020, 1680)h\to KK h$ 
is given by ${\mathcal A}=\Phi_B\otimes H\otimes\Phi_{h}\otimes \Phi_{KK}$~\cite{plb561-258,plb763-29} in the PQCD approach,  according to Feynman diagrams the Fig.~\ref{fig-feyndiag}. 
The hard kernel $H$ contains one hard gluon exchange 
at the leading order in strong coupling $\alpha_s$. The distribution amplitudes $\Phi_B, \Phi_{h}$ and $\Phi_{KK}$ absorb the 
nonperturbative dynamics in the relevant processes. The $\Phi_B$ and $\Phi_{h}$ for $B$ meson and the bachelor final 
state $h$ in this work are the same as those widely employed in the studies of the hadronic $B$ meson decays in the PQCD 
approach, one can find their expressions and parameters in the Appendix of~\cite{jhep2003-162} and the references therein. 

With the subprocesses $\phi \to {K^+ K^-, \bar K^0 K^0}$, and $\phi$ is the $\phi(1020)$ or $\phi(1680)$, the concerned 
quasi-two-body decay amplitudes are given as follows:
\begin{eqnarray}
 {\cal A}(B^+ \to \phi K^+) &=& \frac{G_F} {\sqrt{2}} \big\{V_{ub}^*V_{us}[a_1F^{LL}_{Ah}+C_1M^{LL}_{Ah}]-V_{tb}^*V_{ts}[(a_3+a_4+a_5-\frac{a_7+a_9+a_{10}}{2})F^{LL}_{Th}+(a_4+a_{10})F^{LL}_{Ah}\nonumber\\
 &+&(a_6+a_8)F^{SP}_{Ah}+(C_3+C_4-\frac{C_9}{2}-\frac{C_{10}}{2})M^{LL}_{Th}+(C_5-\frac{C_7}{2})M^{LR}_{Th}+(C_6-\frac{C_8}{2})M^{SP}_{Th}+(C_3\nonumber\\
 &+&C_9) M^{LL}_{Ah}+(C_5+C_7) M^{LR}_{Ah}]\big\} \;, \label{amp1}
\\ 
 {\cal A}(B^+ \to \phi \pi^+) &=& \frac{G_F} {\sqrt{2}}\big\{-V_{tb}^*V_{td}[(a_3+a_5-\frac{a_7+a_9}{2})F^{LL}_{Th}+(C_4-\frac{C_{10}}{2})M^{LL}_{Th}+(C_6-\frac{C_8}{2})M^{SP}_{Th}]\big\} \;, \label{amp2} 
\\ 
 {\cal A}(B^0 \to \phi K^0) &=& \frac{G_F} {\sqrt{2}}
 \big\{-V_{tb}^*V_{td}[(a_3+a_5-\frac{a_7+a_9}{2})F^{LL}_{Th}+(a_4-\frac{a_{10}}{2})(F^{LL}_{Th}+F^{LL}_{Ah})+(a_6-\frac{a_8}{2})F^{SP}_{Ah}+(C_4\nonumber\\
 &-&\frac{C_{10}}{2})M^{LL}_{Th}+(C_3-\frac{C_9}{2})(M^{LL}_{Th}+M^{LL}_{Ah})+(C_5-\frac{C_7}{2})(M^{LR}_{Th}+M^{LR}_{Ah})+(C_6-\frac{C_8}{2})M^{SP}_{Th}]\big\} \;, \label{amp3}
\\ 
 {\cal A}(B^0 \to \phi \pi^0) &=& \frac{-1} {\sqrt{2}} {\cal A}(B^+ \to \phi \pi^+)\;, \label{amp4}
\\ 
 {\cal A}(B_s^0 \to \phi \bar{K}^0) &=& \frac{G_F} {\sqrt{2}}
 \big\{-V_{tb}^*V_{ts}[(a_4-\frac{a_{10}}{2})(F^{LL}_{T\phi}+F^{LL}_{A\phi})+(a_6-\frac{a_8}{2})(F^{SP}_{T\phi}+F^{SP}_{A\phi})+(a_3+a_5-\frac{a_7+a_9}{2})F^{LL}_{Th}\nonumber\\
 &+&(C_4-\frac{C_{10}}{2})M^{LL}_{Th}+(C_6-\frac{C_8}{2})M^{SP}_{Th}+(C_3-\frac{C_9}{2})(M^{LL}_{T\phi}+M^{LL}_{A\phi})+(C_5-\frac{C_7}{2})(M^{LR}_{T\phi}+M^{LR}_{A\phi})]\big\} \;, \quad \label{amp5}
 \\ 
 {\cal A}(B_s^0 \to \phi \pi^0) &=& \frac{G_F} {2}
 \big\{V_{ub}^*V_{us}[a_2 F^{LL}_{T\phi}+C_2 M^{LL}_{T\phi}]-V_{tb}^*V_{ts}[\frac{3}{2}(a_9-a_7)F^{LL}_{T\phi}+\frac{3}{2}C_{10}M^{LL}_{T\phi}+\frac{3}{2}C_8 M^{SP}_{T\phi}]\big\} \;,\label{amp6}
\end{eqnarray}
where $G_F$ is the Fermi coupling constant, $V$'s are the CKM matrix elements. The combinations $a_i$ for the Wilson coefficients 
are defined as
\begin{eqnarray}
  &~& a_1=C_2+\frac{C_1}{3},\quad a_2= C_1+\frac{C_2}{3},\quad a_3= C_3+\frac{C_4}{3},\quad  
          a_4=C_4+\frac{C_3}{3},\quad a_5\,= C_5+\frac{C_6}{3},  \quad \\
  &~& a_6= C_6+\frac{C_5}{3},\quad a_7=C_7+\frac{C_8}{3},\quad a_8= C_8+\frac{C_7}{3},\quad 
          a_9= C_9+\frac{C_{10}}{3},\quad a_{10}= C_{10}+\frac{C_{9}}{3}. \quad 
\end{eqnarray}
It should be understood that the Wilson coefficients $C_i$, the amplitudes $F$ and $M$ for the factorizable and nonfactorizable
Feynman diagrams, respectively, appear in convolutions in momentum fractions and impact parameters $b$.

The amplitudes from Fig.~1 (a) are written as 
\begin{eqnarray}
F^{LL}_{T\phi} &=& 8\pi C_F m^4_B f_h (\zeta-1)\int dx_B dz\int b_B db_B b db \phi_B(x_B,b_B)\big\{\big[(1+z)\phi^0+\sqrt{\zeta}(1-2z)(\phi^s+\phi^t)\big]\nonumber\\
&\times&E_{a12}(t_{a1})h_{a1}(x_B,z,b_B,b)+[\zeta \phi^0+2 \sqrt{\zeta} \phi^s]E_{a12}(t_{a2})h_{a2}(x_B,z,b_B,b) \big\}\;,  
 \label{def-Fll-Tphi}\\
F^{LR}_{T\phi} &=& -F^{LL}_{T\phi}\;, 
\\ 
F^{SP}_{T\phi} &=& 16\pi C_F m^4_B r f_h \int dx_B dz\int b_B db_B b db \phi_B(x_B,b_B)\big\{\big[(\zeta(2z-1)+1)\phi^0+\sqrt{\zeta}((2+z)\phi^s-z\phi^t)\big]\nonumber\\
&\times&E_{a12}(t_{a1})h_{a1}(x_B,z,b_B,b)+\big[x_B\phi^0+2\sqrt{\zeta}(\zeta-x_B+1)\phi^s\big]E_{a12}(t_{a2})h_{a2}(x_B,z,b_B,b)\big\}\;,
\\ 
M^{LL}_{T\phi} &=& 32\pi C_F m^4_B/\sqrt{2N_c} (\zeta-1)\int dx_B dz dx_3\int b_B db_B b_3 db_3\phi_B(x_B,b_B)\phi^A\big\{\big[((1-\zeta)(1-x_3)-x_B\nonumber\\
&-&z\zeta)\phi^0-\sqrt{\zeta}z(\phi^s-\phi^t)\big] E_{a34}(t_{a3})h_{a3}(x_B,z,x_3,b_B,b_3)+\big[(x_3(\zeta-1)+x_B-z)\phi^0+z\sqrt{\zeta}(\phi^s \nonumber\\
&+&\phi^t)\big]E_{a34}(t_{a4})h_{a4}(x_B,z,x_3,b_B,b_3) \big\}\;,\\
M^{LR}_{T\phi} &=& 32\pi C_F r m^4_B/\sqrt{2N_c}\int dx_B dz dx_3\int b_B db_B b_3 db_3\phi_B(x_B,b_B)\big\{\big[((1-x_3)(1-\zeta)-x_B)(\phi^P+\phi^T)\nonumber\\
&\times&(\phi^0+\sqrt{\zeta}(\phi^s-\phi^t))-\sqrt{\zeta} z(\phi^P-\phi^T)(\sqrt{\zeta}\phi^0 -\phi^s-\phi^t)E_{a34}(t_{a3})h_{a3}(x_B,z,x_3,b_B,b_3)\nonumber\\
&+&\big[\sqrt{\zeta}z(\phi^P+\phi^T)(\sqrt{\zeta}\phi^0-\phi^s-\phi^t) +(x_B-x_3(1-\zeta))(\phi^P-\phi^T)(\phi^0+\sqrt{\zeta}(\phi^s-\phi^t))\big]\nonumber\\
&\times&E_{a34}(t_{a4})h_{a4}(x_B,z,x_3,b_B,b_3)\big\}\;,\\
M^{SP}_{T\phi} &=& 32\pi C_F m^4_B/\sqrt{2N_c} (\zeta-1)\int dx_B dz dx_3\int b_B db_B b_3 db_3\phi_B(x_B,b_B)\phi^A\big\{\big[((1-\zeta)(x_3-1)+x_B\nonumber\\
&-&z)\phi^0+\sqrt{\zeta}z(\phi^s+\phi^t)\big]E_{a34}(t_{a3})h_{a3}(x_B,z,x_3,b_B,b_3)+\big[(x_3(1-\zeta)-x_B-z\zeta)\phi^0-z\sqrt{\zeta}(\phi^s\nonumber\\
&-&\phi^t) \big]E_{a34}(t_{a4})h_{a4}(x_B,z,x_3,b_B,b_3) \big\}\;,
\end{eqnarray}
where the color factor $C_F=4/3$ and the ratio $r=m^h_0/m_B$.
The amplitudes from Fig.~1 (b) are written as  
\begin{eqnarray}
F^{LL}_{A\phi} &=& 8\pi C_F m^4_B f_B\int dz dx_3\int b db b_3 db_3 \big\{\big[(1-\zeta)(1-z)\phi^A\phi^0+2r\sqrt{\zeta}\phi^P((z-2)\phi^s-z\phi^t)\big]E_{b12}(t_{b1})\nonumber\\
&\times&h_{b1}(z,x_3,b,b_3)+\big[[(1-x_3)\zeta^2+(2x_3-1)\zeta-x_3]\phi^A\phi^0+2r\sqrt{\zeta}[((1-x_3)\zeta+x_3)(\phi^P+\phi^T)\nonumber\\
&+&(\phi^P-\phi^T)]\phi^s\big]E_{b12}(t_{b2})h_{b2}(z,x_3,b,b_3) \big\}\;,\\
F^{LR}_{A\phi} &=& -F^{LL}_{A\phi}\;,\\
F^{SP}_{A\phi} &=&  16\pi C_F m^4_B f_B \int dz dx_3\int b db b_3 db_3 \big\{\big[2r(1+(z-1)\zeta)\phi^P\phi^0-\sqrt{\zeta}(1-\zeta)(1-z)\phi^A(\phi^s+\phi^t)\big]\nonumber\\
&\times& E_{b12}(t_{b1})h_{b1}(z,x_3,b,b_3)+\left[r\left(x_3(1-\zeta)(\phi^P-\phi^T)-2\zeta \phi^T\right)\phi^0+2\sqrt{\zeta}(\zeta-1)\phi^A\phi^s \right]E_{b12}(t_{b2})\nonumber\\
&\times& h_{b2}(z,x_3,b,b_3)\big\}\;,
\\ 
M^{LL}_{A\phi} &=& 32\pi C_F m^4_B/\sqrt{2N_c} \int dx_B dz dx_3\int b_B db_B b db\phi_B(x_B,b_B)\big\{\big[[(x_3-z-1)\zeta^2+(1+z-2 x_3\nonumber\\
&-&x_B)\zeta+x_3+x_B]\phi^A\phi^0+ r\sqrt\zeta[z(\phi^P-\phi^T)(\phi^s+\phi^t)+((1-x_3)(1-\zeta)-x_B)(\phi^P+\phi^T)(\phi^s\nonumber\\
&-&\phi^t)-4\phi^P\phi^s]\big] E_{b34}(t_{b3})h_{b3}(x_B,z,x_3,b_B,b)+\big[(1-\zeta)^2(z-1)\phi^A\phi^0+ r\sqrt\zeta [(\zeta(1-x_3)+x_3\nonumber\\
&-&x_B)(\phi^P-\phi^T)(\phi^s+\phi^t)+ (1-z)(\phi^P+\phi^T)(\phi^s-\phi^t)]\big]E_{b34}(t_{b4})h_{b4}(x_B,z,x_3,b_B,b)\big\}\;,\\
M^{LR}_{A\phi} &=& 32\pi C_F m^4_B/\sqrt{2N_c} \int dx_B dz dx_3\int b_B db_B b db \phi_B(x_B,b_B)\big\{\big[r [(2+\zeta x_3-x_3-x_B)(\phi^P+\phi^T)\nonumber\\
&-&\zeta z(\phi^P-\phi^T)-2 \zeta \phi^P]\phi^0+\sqrt\zeta(1-\zeta)(1+z)\phi^A(\phi^s-\phi^t)\big]E_{b34}(t_{b3})h_{b3}(x_B,z,x_3,b_B,b)\nonumber\\
&+& \big[r[(x_3(1-\zeta)-x_B)(\phi^P+\phi^T)+\zeta z(\phi^P-\phi^T)+2\zeta \phi^T]\phi^0+\sqrt\zeta(1-\zeta)(1-z)\phi^A(\phi^s-\phi^t) \big]\nonumber\\
&\times&E_{b34}(t_{b4})h_{b4}(x_B,z,x_3,b_B,b)\big\}\;,
\end{eqnarray}
\begin{eqnarray}
M^{SP}_{A\phi} &=& 32\pi C_F m^4_B/\sqrt{2N_c} \int dx_B dz dx_3\int b_B db_B b db\phi_B(x_B,b_B)\big\{\big[(\zeta-1)[(\zeta-1)z+1]\phi^A\phi^0+ r\sqrt\zeta\nonumber\\
&\times& [((1-\zeta)(x_3-1)+x_B)(\phi^P-\phi^T)(\phi^s+\phi^t)-z(\phi^P+\phi^T)(\phi^s-\phi^t)+4\phi^P\phi^s]\big]\nonumber\\
&\times&E_{b34}(t_{b3})h_{b3}(x_B,z,x_3,b_B,b)+ \big[[(\zeta-1)(x_3(\zeta-1)+x_B)+\zeta z(1-\zeta)]\phi^A\phi^0+r\sqrt\zeta [(z-1)\nonumber\\
&\times&(\phi^P-\phi^T)(\phi^s+\phi^t)+((\zeta-1)x_3+x_B-\zeta)(\phi^P+\phi^T)(\phi^s-\phi^t)]\big] E_{b34}(t_{b4})h_{b4}(x_B,z,x_3,b_B,b)\big\}\;.
\end{eqnarray}
The amplitudes from Fig.~1 (c) are written as  
\begin{eqnarray}
F^{LL}_{Th} &=& 8\pi C_F m^4_B F_K\int dx_B dx_3\int b_B db_B b_3 db_3 \phi_B(x_B,b_B)\big\{\big[(1-\zeta)[(x_3(\zeta-1)-1)\phi^A+ r(2x_3-1)\phi^P]-r(1 \quad  \nonumber\\
&+&\zeta-2x_3(1-\zeta))\phi^T\big]E_{c12}(t_{c1})h_{c1}(x_B,x_3,b_B,b_3)+\left[x_B(1-\zeta)\zeta\phi^A-2r(1-\zeta(1-x_B))\phi^P\right]E_{c12}(t_{c2}) \nonumber\\
&\times& h_{c2}(x_B,x_3,b_B,b_3)\big\}\;,\\
F^{LR}_{Th} &=& F^{LL}_{Th},
\\  
M^{LL}_{Th} &=& 32\pi C_F m^4_B/\sqrt{2N_c} \int dx_B dz dx_3\int b_B db_B b db
  \phi_B(x_B,b_B)\phi^0 \big\{\big[(x_B+z-1)(1-\zeta)^2\phi^A +r[\zeta(x_B+z)\nonumber\\
&\times&(\phi^P+\phi^T) +x_3(1-\zeta)(\phi^P-\phi^T)-2\zeta\phi^T]\big] E_{c34}(t_{c3})h_{c3}(x_B,z,x_3,b_B,b)+\big[(\zeta-1)[x_3(\zeta-1)+x_B\nonumber\\
&-&z]\phi^A+r[x_3(\zeta-1)(\phi^P+\phi^T)-(x_B-z)\zeta(\phi^P_K-\phi^T_K)]\big]E_{c34}(t_{c4})h_{c4}(x_B,z,x_3,b_B,b)\big\}\;,\\
M^{LR}_{Th} &=& 32\pi C_F m^4_B\sqrt{\zeta}/\sqrt{2N_c} \int dx_B dz dx_3\int b_B db_B b db \phi_B(x_B,b_B)\big\{\big[(1-x_B-z)(\zeta-1)(\phi^s+\phi^t)\phi^A-r\nonumber\\
&\times&(x_3(1-\zeta)+\zeta)(\phi^s-\phi^t)(\phi^P+\phi^T)-r(1-x_B-z)(\phi^s+\phi^t)(\phi^P-\phi^T)\big]E_{c34}(t_{c3})h_{c3}(x_B,z,x_3,b_B,b)\nonumber\\
&+&\big[(z-x_B)(1-\zeta)(\phi^s-\phi^t)\phi^A+rx_3(1-\zeta)(\phi^s+\phi^t)(\phi^P+\phi^T)+r(z-x_B)(\phi^s-\phi^t)(\phi^P-\phi^T)\big]\nonumber\\
&\times&E_{c34}(t_{c4})h_{c4}(x_B,z,x_3,b_B,b)\big\}\;,\\
M^{SP}_{Th} &=& 32\pi C_F m^4_B/\sqrt{2N_c} \int dx_B dz dx_3\int b_B db_B b db \phi_B(x_B,b_B)\phi^0\big\{\big[(\zeta(x_3-1)-x_3+x_B+z-1)(1-\zeta)\phi^A\nonumber\\
&+& r x_3(1-\zeta)(\phi^P+\phi^T)+ r\zeta(x_B+z)(\phi^P-\phi^T)+ 2r\zeta\phi^T\big] E_{c34}(t_{c3})h_{c3}(x_B,z,x_3,b_B,b)+\big[(z-x_B) \nonumber\\
&\times&(1-\zeta)^2\phi^A+r\zeta(z-x_B)(\phi^P+\phi^T)-rx_3(1-\zeta)(\phi^P-\phi^T)\big]E_{c34}(t_{c4})h_{c4}(x_B,z,x_3,b_B,b)\big\}\;.
\end{eqnarray}
The amplitudes from Fig.~1(d) are written as  
\begin{eqnarray}
F^{LL}_{Ah} &=& 8\pi C_F m^4_B f_B \int dz dx_3\int b db b_3 db_3\big\{\big [(x_3(1-\zeta)-1)(\zeta-1)\phi^0\phi^A+ 2r\sqrt\zeta\phi^s[x_3(\zeta-1)(\phi^P-\phi^T) +2\phi^P]\big]\nonumber\\
&\times&E_{d12}(t_{d1})h_{d1}(z,x_3,b,b_3)+\big[z(\zeta-1)\phi^0\phi^A-2r\sqrt\zeta[z(\phi^s+\phi^t)+(1-\zeta)(\phi^s-\phi^t)]\phi^P\big] E_{d12}(t_{d2})\nonumber\\
&\times&h_{d2}(z,x_3,b,b_3) \big\}\;,\\
F^{LR}_{Ah} &=& -F^{LL}_{Ah}\;,\\
F^{SP}_{Ah} &=& 16\pi C_F m^4_B f_B  \int dz dx_3\int b db b_3 db_3 \big\{\big[(\zeta-1)[2\sqrt\zeta\phi^s\phi^A+r(1-x_3)\phi^0 \phi^P]- r[\zeta+x_3(\zeta-1)+1]\phi^0\phi^T\big]\nonumber\\
&\times& E_{d12}(t_{d1})h_{d1}(z,x_3,b,b_3)+ \big[z\sqrt\zeta(\zeta-1)(\phi^s-\phi^t)\phi^A+2r(z \zeta+\zeta-1)\phi^0\phi^P \big]E_{d12}(t_{d2})h_{d2}(z,x_3,b,b_3)\big\}\;,\quad
\\ 
M^{LL}_{Ah} &=& 32\pi C_F m^4_B/\sqrt{2N_c} \int dx_B dz dx_3\int b_B db_B b db\phi_B(x_B,b_B) \big\{\big[[(x_B+z-1)\zeta^2+ (1-2x_B- 2z)\zeta+x_B\nonumber\\
&+&z]\phi^0\phi^A- r\sqrt\zeta[(\eta(1-x_3)+x_3)(\phi^s-\phi^t)(\phi^P+\phi^T)+ (1-x_B-z)(\phi^s+\phi^t)(\phi^P-\phi^T)- 4\phi^s\phi^P]\big]\nonumber\\
&\times&E_{d34}(t_{d3})h_{d3}(x_B,z,x_3,b_B,b)+ \big[(\zeta-1)((1-x_3)(1-\zeta)+\zeta(x_B-z))\phi^0\phi^A+r\sqrt\zeta[(x_B-z)(\phi^s-\phi^t)\nonumber\\
&\times& (\phi^P+\phi^T)+(1-\zeta)(x_3-1)(\phi^s+\phi^t)(\phi^P-\phi^T)] \big]E_{d34}(t_{d4})h_{d4}(x_B,z,x_3,b_B,b) \big\}\;,\\
M^{LR}_{Ah} &=&  32\pi C_F m^4_B/\sqrt{2N_c} \int dx_B dz dx_3\int b_B db_B b db \phi_B(x_B,b_B)\big\{\big[\sqrt\zeta(1-\zeta)(x_B+z-2)(\phi^s+\phi^t)\phi^A\nonumber\\
&+& r\phi^0[\zeta (x_B+z-1)(\phi^P+\phi^T)+ (1+x_3-\zeta x_3)(\phi^P-\phi^T)-2\zeta \phi^T]\big]E_{d34}(t_{d3})h_{d3}(x_B,z,x_3,b_B,b)\nonumber\\
&+&\big[\sqrt\zeta(1-\zeta)(x_B-z)(\phi^s+\phi^t)\phi^A+r\phi^0[\zeta(x_B-z)(\phi^P+\phi^T)+(1-\zeta)(1-x_3)(\phi^P-\phi^T)]\big]\nonumber\\
&\times&E_{d34}(t_{d4})h_{d4}(x_B,z,x_3,b_B,b)\big\}\;,
\end{eqnarray}
\begin{eqnarray}
M^{SP}_{Ah} &=& 32\pi C_F m^4_B/\sqrt{2N_c} \int dx_B dz dx_3\int b_B db_B b db\phi_B(x_B,b_B)\big\{\big[(\zeta-1)[x_3(\zeta-1)-\zeta(x_B+z)+1]\phi^0\phi^A\nonumber\\
&-& r\sqrt\zeta[(x_B+z-1)(\phi^s-\phi^t)(\phi^P+\phi^T)+ (\zeta x_3-\zeta-x_3)(\phi^s+\phi^t)(\phi^P-\phi^T)+4\phi^s\phi^P]\big]E_{d34}(t_{d3})\nonumber\\
&\times&h_{d3}(x_B,z,x_3,b_B,b)+ \big[(\zeta-1)^2(z-x_B)\phi^0\phi^A- r\sqrt\zeta[(1-\zeta)(x_3-1)(\phi^s-\phi^t)(\phi^P+\phi^T)+ (x_B-z)\nonumber\\
&\times&(\phi^s+\phi^t)(\phi^P-\phi^T) \big] E_{d34}(t_{d4})h_{d4}(x_B,z,x_3,b_B,b) \big\}\;. \label{def-Msp-ah}
\end{eqnarray}

For the errors induced by the parameter ${\mathcal P}\pm\Delta{\mathcal P}$ for the $\mathcal{B}$ and ${\mathcal A}_{CP}$ 
in the numerical calculation of this work, we employ the formulas~\cite{jhep2003-162}
\begin{eqnarray}
\Delta\mathcal{B}=\left|\frac{\partial \mathcal{B}}{\partial{\mathcal P}}\right|\Delta{\mathcal P},\qquad
\Delta{\mathcal A}_{CP}=\frac{2(\mathcal{B}\Delta\overline{\mathcal{B}}-\overline{\mathcal{B}}\Delta\mathcal{B})}
{(\overline{\mathcal{B}}+\mathcal{B})^2}.
\end{eqnarray}
The PQCD functions which appear in the factorization formulas, the Eqs.~(\ref{def-Fll-Tphi})-(\ref{def-Msp-ah}), can be found in 
the Appendix B of~\cite{jhep2003-162}.


\end{document}